\documentclass[final,5p,times,twocolumn]{elsarticle}

\usepackage{graphicx}
\usepackage{array}
\usepackage{dblfloatfix}
\journal{Nuclear Instruments and Methods in Physics Research A}
\usepackage{amssymb}
\usepackage{siunitx}
\usepackage{booktabs}
\usepackage{IEEEtrantools}
\usepackage{amsmath}
	\numberwithin{equation}{section}
\usepackage{hyperref}
\usepackage{cleveref}
\usepackage{subfigure}
\usepackage[skip=3pt]{caption}
\usepackage{siunitx}

\usepackage[figuresright]{rotating}

\begin{document}

\begin{frontmatter}


%%\title{Title\tnoteref{label1}}

\title{EUPRAXIA@SPARC\_LAB: \\ Beam Dynamics studies for the X-band Linac}

\author[lnf]{C. Vaccarezza\corref{mycorrespondingauthor}}
\cortext[mycorrespondingauthor]{Corresponding author}
\ead{\string\href{mailto:Cristina.Vaccarezza@lnf.infn.it}{cristina.vaccarezza@lnf.infn.it}}
\author[lnf]{D. Alesini}
\author[mi]{A. Bacci}
\author[Tov]{A. Cianchi}
\author[lnf]{E. Chiadroni}
\author[lnf]{M. Croia}
\author[lnf,Sap]{M. Diomede}
\author[lnf]{M. Ferrario}
\author[lnf]{A. Gallo}
\author[lnf]{A. Giribono}
\author[cern]{A. Latina}
\author[lnf]{A. Marocchino}
\author[mi]{V. Petrillo}
\author[lnf]{R. Pompili}
\author[lnf]{S. Romeo}
\author[mi]{M. Rossetti Conti}
\author[mi]{A.R. Rossi}
\author[mi]{L. Serafini}
\author[lnf]{B. Spataro}

\address[lnf]{INFN-LNF, Via E. Fermi 40, Frascati, Italy}
\address[mi]{INFN-MI, Via E. Celoria 16, Milano, Italy}
\address[Tov]{University of Rome Tor Vergata and INFN, via delle Ricerca Scientifica,Roma, Italy}
\address[Sap]{Sapienza University, Piazzale Aldo Moro 5, Roma, Italy}
\address[cern]{CERN, CH-1211 Geneva-23, Switzerland}
%\address[sapienza]{Sapienza University, Piazzale Aldo Moro 5, Rome, Italy}
\begin{abstract}
In the framework of the Eupraxia Design Study an advanced accelerator facility EUPRAXIA@SPARC\_LAB has been proposed to be realized at Frascati (Italy) Laboratories of INFN. Two advanced acceleration schemes will be applied, namely an ultimate high gradient 1 GeV X-band linac together with a plasma acceleration stage to provide accelerating gradients of the GeV/m order.
A FEL scheme is foreseen to produce X-ray beams within 3-10 nm range. A 500-TW Laser system is also foreseen for electron and ion production experiments and a Compton backscattering Interaction is planned together with extraction beamlines at intermediate electron beam energy for neutron beams and THz radiation production. The electron beam dynamics studies in the linac are here presented together with the preliminary machine layout.
\end{abstract}

\begin{keyword}

%% PACS codes here, in the form: 
%%\PACS code \sep code


\end{keyword}

\end{frontmatter}

%%\linenumbers

\section{Introduction}
Within the Horizon 2020 program the EuPRAXIA Design Study (European Plasma Research Accelerator with eXcellence In Applications) \cite{Eupraxia_EU} will propose in 2019 the first European Research Infrastructure devoted to demonstrate the application of plasma acceleration to generate high brightness beams with energy between 1-5 GeV. In this framework the EuPRAXIA@SPARC\_LAB project \cite{Eupraxia_SPARC_LAB} is meant to realize at the INFN-LNF Laboratories (Frascati, Italy) a machine resulting a unique combination of a high brightness GeV-range electron beam generated in a state-of-the-art advanced compact linac, a 0.5 PW-class laser system and the first 5th generation light source, aiming to candidate the LNF as host for the EuPRAXIA european facility.
A X-band RF Linac has been designed to couple a SPARC-like high brightness photoinjector \cite{SPARC_proj} with a short wavelength FEL ($\lambda_r= 3 nm$) and a Compton backscattering radiation source with energy up to $E=1 GeV$. To test the robustness and flexibility of the lattice design three main working points have been studied  and here described in terms of the 6D phase space of the electron beam at the Linac exit that means at the plasma capillary entrance (for WP1-WP2) where the required beam transverse size is around $\sigma_{x,y} \approx 1-4 \mu m$ according to the matching condition for the acceleration in the plasma, and at the undulator entrance (WP3, conventional linac only) where a transverse beam size $\sigma_{x,y} \approx 30-60 \mu m$ is required to match the undulator lattice. 
This choice represents a first test of the linac design for possible machine configurations depending on the electron beam characteristics obtained at the exit of the foreseen SPARC-like Photoinjector (briefly described in chapter 2 and extensively in ref. \cite{SPARC_proj, Anna_PI}) : a low charge/high current electron beam, WP1, where the full RF longitudinal compression takes place in the photoinjector (rms bunch length $\sigma_z\approx 6 \mu m$, $E=100 MeV$), suitable for acceleration in the plasma stage at the energy of $E\approx500 MeV $ after the X-band Linac; a high charge/low current, WP3, with the combination of RF compression in the photoinjector and final longitudinal compression in the magnetic chicane from a rms bunch length of $\sigma_z\approx110 \mu m$ at the initial energy of  $E=170 MeV$ down to $\sigma_z\approx10-20 \mu m$, final energy $E\approx1 GeV$, to check the contribution of the Coherent Synchrotron Radiation effect to the transverse emittance dilution in the magnetic chicane; plus a low charge/low current electron beam, WP2, to test the hybrid longitudinal compression scheme from an initial rms bunch length of $\sigma_z\approx20 \mu m$, initial energy  $E=170 MeV$, down to  $\sigma_z\approx4 \mu m$ at a final energy $E\approx500 MeV$ on a beam again to be injected  in the plasma acceleration stage. The Compton Source requirements for the electron beam fall among the previous three working points characteristics and the corresponding working point will not be described here. Finally the choice of X-band ($f_{rf}=11.994 GHz$) as the RF frequency for the linac accelerating structures makes this a frontier project and puts among the main themes the machine sensitivity to the active elements misalignments; in this paper the first results of simulation studies are presented.
\section{Machine Layout}
The EuPRAXIA@SPARC\_LAB accelerator is approximately 50 m long, the electron beam is generated in a twelve meters long high brightness SPARC-like S-band Photoinjector described in ref. \cite{SPARC_proj}:  1.6 cell S-band RF gun where a Cu photocathode is mounted and driven by a 50 ${\mu J}$ Ti:Sapphire laser with a four coils solenoid for the emittance compensation; three TW SLAC type S-band linac sections follow for a final energy ranging between  $E=100-230 MeV$ depending on the applied RF compression factor as described in \cite{SPARC_proj,Anna_PI}.  The downstream X-band RF Linac can bring the electron beam energy up to $E_{max}\approx 1GeV$; at the X-band Linac exit a Plasma Acceleration Structure (PLAS) is foreseen and after this two separate transfer lines deliver the electron beam to the Undulator and to the Compton Interaction point respectively, see schematic layout in Fig. 1. 
\begin{figure*}[h!]
\begin{center}
\includegraphics[width=0.8\textwidth]{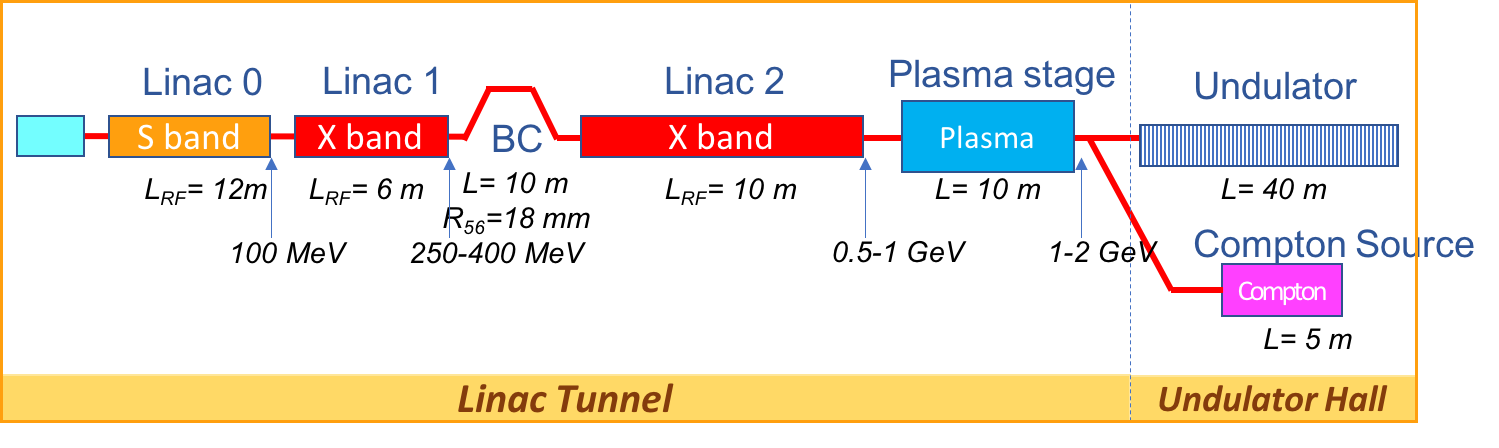}
%\epsfile{Linac_layout,scale=0.8}
\caption{Electron beam acceleration schematic layout. In this lego scheme the main structures lengths are reported together with the RF active length ($L_{RF}$), and the "Plasma Interaction" box embeds also the injection/extraction beamline.}
\label{fig:schematic_layout}
\end{center}
\end{figure*}
A four dipole magnetic chicane, 10 m long, is inserted in the X-band Linac between the two sections Linac1(L1) and Linac2 (L2), for longitudinal compression and phase space manipulation of the electron beam.
In order to satisfy the requirements of the SASE-FEL radiation source foreseen in the EuPRAXIA@SPARC\_LAB project a FWHM bunch current of 2-3 kA must be delivered to entrance of the undulator at the energy of 1 GeV, with a very good collimation in the 6D phase space, see table \ref{table1}. 
\begin{table}[h!]
%  \begin{center}
    \caption{Electron beam parameters for plasma driven/conventional FEL at EuPRAXIA@SPARC\_LAB.}
    \label{table1}
    \begin{tabular}{l|c|c|c} % <-- Alignments: 1st column left, 2nd middle and 3rd right, with vertical lines in between
    \textbf &&{1 GeV} &{1 GeV} \\
    \textbf &{Units} &{Plasma driven} &{X-band only} \\
      \hline
      Bunch charge& $pC$&30 & 200\\
      Bunch length rms & $fs$&10 & 60 \\
      Peak Current rms & $kA$&3 & 2\\
      Rms Energy Spread& \%&1. & 0.1 \\
      Slice Energy Spread& \%&0.1 & 0.05 \\
      Rms norm. emittance & $\mu m$&1. & 1. \\
      Slice nor. emittance & $\mu m$&$< 1. $& $< 1.$ \\
      Slice Length &$ \mu $m&0.75 & 1. \\
      Radiation wavelength & $nm$&3 & 3 \\
      $\rho $ & $\times 10^{-3}$&$< 1$ &$< 1$ \\
      Undulator period & $cm$&1.5 & 1.5 \\
      K & &0.987 & 0.987 \\
     \end{tabular}
%  \end{center}
\end{table}
The 1 GeV energy can be achieved by means of a single stage of plasma acceleration, few centimeters long, coupled with the RF Linac operating at 500 MeV, or with the conventional operation of the Linac at twice the accelerating gradient in the X-band sections. 
%%%%%%%%-----------%%%%%%%%%
The goal of the project is to operate plasma acceleration at approximately $10^{16}cm^{-3}$, a plasma
density that can be used to produce electric fields of 1-2 GV/m and characterized by a plasma wavelength of $\lambda_p \approx$ \SI{300}{\micro\metre}
that allows for realistic bunch separation with the use of a COMB technique \cite{ChiadroniCuba2017,Anna_PI}. Such accelerating gradients are tailored for our specific 
envisioned experiment \cite{beam_driven,stefanoromeo}, where the foreseen parameters will allow for good beam loading compensation and lower quality
depletion. This matches with the chosen  plasma input energy of 500 MeV, highly rigid bunch, that limits transverse bunch evolution and the 
consequent transverse emittance dilution within the plasma.
%%%%%%%%-----------%%%%%%%%%At the same time the conventional 1 GeV RF Linac is able to provide the high photon flux operation both for the FEL, same wavelength, and for the Compton Source.
Three working points (WP) have been studied up to now for the accelerator according to the beam characteristic at the exit of the previously described photoinjector \cite {Anna_PI}:
\begin{enumerate}
\item WP1: Low Charge-High Current: 30 pC-3KA (FWHM) per bunch, initial energy $E_{in}\approx 100 MeV$, suitable both for Beam Driven and Laser driven acceleration in Plasma at the X-band Linac exit with final energy $E_{fin}\approx 500 MeV$,
\item WP2: Low Charge-Low Current: 30 pC-100A per bunch, $E_{in}\approx170 MeV$, coupled with a magnetic longitudinal compression stage ($R_{56}=9 mm$), in the foreseen C-shape magnetic chicane, to reach the desired current $I_{pk} = 3 kA$ (Hybrid scheme), suitable both for Beam Driven and Laser driven acceleration in Plasma at $E_{fin}\approx 500 MeV$,
\item WP3: High charge-Low Current: 200 pC-70 A, $E_{in}=170 MeV$-$E_{fin}\approx 1 GeV$, with and without the longitudinal bunch compression in the magnetic chicane ($R_{56}=16 mm$) to serve the SASE-FEL, with peak current $I_{pk}=2kA$ , and the Compton and THz sources in the high flux operation scheme.
\end{enumerate} 
For these three working points start-to-end simulations of the electron beam dynamics have been performed on the nominal cases and reported in ref. \cite{Anna_PI,Andrea_LW} for the photoinjector and the plasma acceleration stage respectively, while the results for the FEL radiation are reported in \cite{Vittoria}, here the transport in the X-band Linac is described looking at the machine sensitivity to misalignments and jitters.
\subsection{The X-band Linac}
The Eupraxia@SPARCLAB X-band linac consists of two sections L1 and L2 located before and after the magnetic chicane respectively. Twelve X-band accelerating sections, 50 cm long, are foreseen for L1 and twenty for L2. According to the RF power system design \cite{M_Diomede} the maximum accelerating gradient applied is $E_{acc}\approx 60 MV/m$ through all L1 and L2, to reach the required energy and energy spread for the electron beam in the conventional RF operation scheme.  An increased power configuration can be also implemented progressively in a machine upgrade plan to provide overhead and flexibility to the operation, and ultimately to reach higher beam energies with the accelerating gradient raised up to $\approx 80 MV/m$.
Downstream the L2 linac a 10 m long matching section is foreseen to inject the electron beam in the plasma stage, (PLAS), $10-30$ centimeters long, and to capture and match it from the capillary exit to the undulator entrance. 
A four dipole magnetic chicane, 10 m long, is foreseen for phase space manipulation and longitudinal compression of the bunch, (WP2 and WP3); at the same time when the chicane dipoles are switched off, the foreseen straight beamline accomodates the middle energy diagnostic station for beam parameters measurement. Finally, at the PLAS exit a dogleg bend system (DL), parallel to the undulator, is foreseen for energy and energy spread analysis, transverse emittance measurement, and final beam transport to the Compton Source Interaction Point. 
Parameters, design criteria, and beam dynamics results  are discussed in the following paragraphs. 
\subsection{Longitudinal Beam Dynamics}
\begin{figure}[h!]
\begin{center}
\includegraphics[width=0.45 \textwidth]{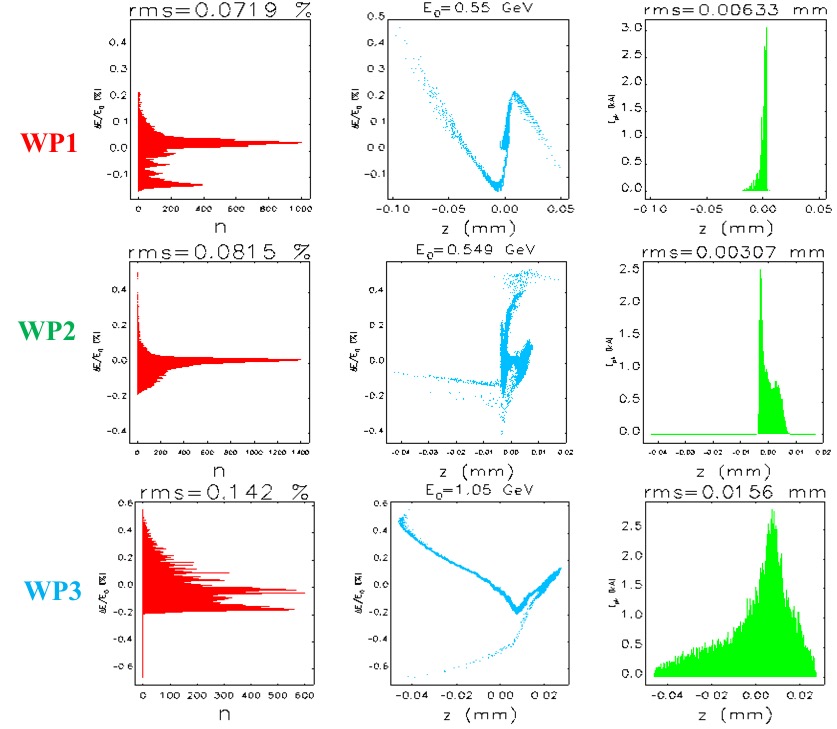}
%\epsfile{Linac_layout,scale=0.8}
\caption{Longitudinal phase space of WP1-2 case at the PAS capillary entrance, and at the undulator entrance for WP3 (Elegant code).}
\label{fig:Long_pspace}
\end{center}
\end{figure}

The two Linac sections L1 and L2 have been optimized to provide the required beam acceptance, from photoinjector and after the magnetic chicane, for the three considered WP's, and the best focusing strength for the lattice has been found with a betatron phase advance per cell of $15^{\circ}$ and $28^{\circ}$ for L1 and L2 respectively.  
In order to reach the desired peak current for WP2 and WP3 a magnetic compression, with $R_{56} = 9\div16 mm$, is applied by means of the C-shape chicane inserted between L1 and L2 linac sections. In table \ref{L1_L2_table} the L1 and L2 main parameter list is reported.
%%\begin{table}[h!]
\begin{table*}[h!]
  \begin{center}
    \caption{L1 and L2 Linacs parameter list.}
    \label{L1_L2_table}
    \begin{tabular}{l|c|ccc|ccc|} % <-- Alignments: 1st column left, 2nd middle and 3rd right, with vertical lines in between
%   {Beam Parameter} &{Unit}& \multicolumn{3}{c}{L1}&\multicolumn{3}{c}{L2}& \\
{Beam Parameter} &{Unit}& &{L1}&&&{L2}& \\
      \hline
&&WP1&WP2&WP3&WP1&WP2&WP3\\
Initial energy (Photoinjector exit) &$GeV$&0.100&0.170&0.170&0.210&0.284&0.505\\
Final energy&$GeV$&0.210&0.284&0.550&0.550&0.550&1.060\\
Active Linac length&$m$&&6.0&&&10.0&\\
Acc. Gradient&$MV/m$&20.0&20.0&57.0&36.0&26.8&57.0\\
RF phase (0 crest)&$deg$&-20.0&-20.0&-12.0&-19.5&0&+15.0\\
Initial energy spread&$\%$&0.30&0.22&0.67&0.15&0.22&0.47\\
Final energy spread&$\%$&0.15&0.22&0.47&0.07&0.06&0.09\\
Bunch length&$mm$&0.006&0.020&0.112&0.006&0.004&0.020\\
     \end{tabular}
  \end{center}
\end{table*}
For each WP 30k macroparticles have first simulated in the photoinjector by means of the TSTEP code ~\cite{Tstep,Anna_PI} and then tracked trough the linac using the Elegant code ~\cite{Elegant}, where the considered asymptotic values of the longitudinal and transverse wake functions have been calculated according to  \cite{K_Bane}:
\begin{equation}
W_{0\parallel}\approx \frac{Z_0c}{\pi a^2}exp\biggl( -\sqrt{\frac{s}{s_1}} \biggr) (V/Cm)
\end{equation}
\begin{equation}
W_{0\perp}\approx \frac{4Z_0cs_2}{\pi a^4}\biggl [ 1-\biggl (1+sqrt{\frac{s}{s_1}} \biggr ] exp\biggl( -\sqrt{\frac{s}{s_2}} \biggr) (V/Cm^2)
\end{equation}
\begin{equation}
s_1=0.41\frac{a^{1.8}g^{1.6}}{L^{2.4}},s_2=0.17\frac{a^{1.79}g^{.38}}{L^{1.17}}
\end{equation}
where $Z_0$ is the free space impedance, $c$ is the light velocity, $a=3.2 mm$ is the iris radius, $L=8.332 mm$ is the cell length and $g=6.495 mm$ is the cavity length for the pill box model representing our X-band structure ~\cite{M_Diomede}
The obtained results are shown in figure \ref{fig:Long_pspace}, where the energy spread and the distribution of the energy and current along the electron bunch is shown as obtained with the simulation at the linac exit, in agreement with the table \ref{table1} requirements.
 \subsection{Transverse Beam Dynamics}
The three considered working points mainly differ for the compression factor applied in the velocity bunching scheme at the photoinjector and for the momentum compaction value of the magnetic chicane; moreover the WP1-WP2 working points require a strong final focusing at the PLAS entrance and a proper capture section after the acceleration, more demanding than the matching optics needed to inject the beam coming from the Linac directly into the undulator. 
\subsubsection{WP 1-2 (30pC,3kA,500 MeV)}
These two working points are meant to provide an electron beam for injection in the PLAS capillary, $Q=30pC$, $I=3kA(FWHM)$, with a $1-4 \mu m$ of transverse spot-size and less than 0.1 \% of energy spread as reported in Table \ref{table1}. For the WP1 case the electron bunch is fully compressed in the photoinector by means of the velocity bunching operation scheme, while for the WP2 case the final compression occurs in the magnetic chicane (mostly aiming to test the CSR radiation contribution to the emittance dilution in case of beam phase space manipulation in the chicane). Similar results are obtained in both cases. Before entering the plasma capillary a focusing triplet of permanent quadrupoles is foreseen at a distance of few centimeters from the plasma entrance together with a longer FODO arrange at the exit to capture a beam with typical $\beta$-function $\beta_{x,y} \approx1-5 mm$. The gradient of the first three permanent quadrupoles is around $G\approx300 T/m$ with a magnetic length of $5-10 cm $ while at the exit of the plasma, an array of ten alternating quadrupoles, 8.5 cm long, is foreseen with same gradient. 
%This, still preliminary, optimization of the focusing/extraction system tends to minimize the emittance dilution due to chromatic effects. 
A longitudinal position adjustment setup to tune the strength of the final focus array and latest generation of tunable permanent quadrupoles are under study to increase as much as possible the tunability of the magnet arrange, mostly at the plasma exit, and widen the energy acceptance of the transferline. 
%In Fig. \ref{fig:cap_pspace} the transverse phase space of the electron beam is shown at the capillary entrance. 
%\begin{figure}[h!]
%\begin{center}
%includegraphics[width=0.42\textwidth]{cap_pspace.png}
%\epsfile{Linac_layout,scale=0.8}
%\caption{Transverse spotsize and transverse-longitudinal distribution of WP1 case electron beam at the plasma capillary entrance, as obtained with the Elegant code simulation}
%\label{fig:cap_pspace}
%\end{center}
%\end{figure
\subsubsection{WP 3 (200pC,2kA,1 GeV}
The WP3 case copes with the emittance dilution due to the CSR effect occurring in the magnetic chicane kept as short as possible due to room availability, it results in a final projected emittance dilution of about 60\%, going from the initial $\epsilon_{n,x,y}=0.5 \mu rad$ to the final $\epsilon_{n,x,y}=0.8 mm$ $mrad$, nevertheless the beam quality results to be preserved in the slices corresponding to the highest current of the electron bunch, see \cite{Vittoria}.
\section{Misalignments and jitters effect in the X-band Linac}
\begin{figure}[h!]
\begin{center}
\includegraphics[width=0.50\textwidth]{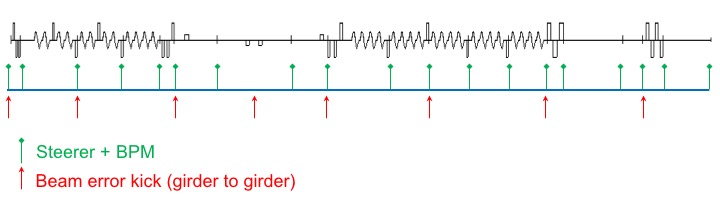}
%\epsfile{Linac_layout,scale=0.8}
\caption{BPM and Steerers considered distribution along the X-band Linac (green arrows), plus the additional girder to girder misalignment errors (red arrows) .}
\label{fig:BPM_Steerer}
\end{center}
\end{figure}

The described Linac lattice represents the first iteration of a more in-depth optimization work to be performed in the very next, but it provides the baseline to address the misalignment and jitter effects on the electron beam quality at the linac exit.
For this reason a preliminary Beam Position Monitors (BPM) and steerer magnets distribution has been considered  as reported in figure \ref{fig:BPM_Steerer}, and static and dynamic errors of the active elements have been randomly simulated to check the robustness and feasibility of the Linac design; the BPM error was no considered at this point to isolate the contribution of the RF sections and quadrupoles misalignment.
The sensitivity of the Linac to misalignments and jitters has been studied in terms of the electron beam centroid distribution, beam spot-size, emittance and energy spread at the capillary entrance (WP1-WP2) or at the undulator entrance (WP3), while along the Linac the minimum-maximum value of the trajectories (centroid) and  the beam transverse envelope have been considered to evaluate the required electron beam stay-clear in the X-band sections.
The misalignments/jitters study has been performed for each working point adding at first alignment errors (static) on each 50 cm long  X-band structure and quadrupole magnet: the applied errors have been randomly generated inside a $\Delta_{x,y}=70 \mu m$ for RF structures and quadrupole magnets, plus an additional misalignment  $\Delta_{x,y}=150 \mu m$ representing girder to girder transition (red arrows on fig. \ref{fig:BPM_Steerer}). 100 simulation runs have been performed with the Elegant code applying also the proper trajectory correction with the steerer magnets. For each of this static error and corrected trajectory simulation other 100 runs have been performed adding a random 0.1\% jitter (dynamic error) to the steerer and quadrupole strength.  A total of 10000 runs have been done in this way for each WP and the results are reported in table \ref{summary}. At this preliminary stage of the Linac sensitivity study the resulting stability of the centroid position can be considered acceptable resulting as a maximum the 40\% of the rms beam size which is the case of the horizontal plane for the WP1 working point, the same comment can applied to the 10\% deviation in the horizontal beam size for case WP2. More severe are the deviations on the horizontal plane for the WP3 case and this will be evaluated with simulations of FEL radiation in presence of such machine errors.
The results obtained for the rms beam size envelope along the Linac show a maximum value of $\Delta_{\sigma_{x,y}}\pm 200 \mu m$ while the obtained
trajectory envelopes show for the centroid positions a maximum oscillation of $\Delta_{C_{x,y}}\pm 500 \mu m$ for the WP1-WP3 cases, indicating that a beam stay-clear of more than $\pm 4$ rms beam size is possible even in the x-band cells with iris radius $a\approx 2.4 mm$. Only for WP2 at the entrance of Linac1 the trajectory envelope reaches $\pm 900\mu m$ and this will be investigated and fixed. 
It has to be noticed that no dispersion free steering has been applied up to now and that different steerer location together with the calculated kicks distribution will be studied in the future to check the best arrangement of the BPM-Steerer set.
\begin{table}[h!]
  \begin{center}
    \caption{Summary table of the static-dynamic errors study for the X-band Linac working points: WP1-2 columns refer at capillary entrance, WP3 at undulator entrance.}
    \label{summary}
    \begin{tabular}{p{2.8 cm}l| c| c| c| c|} % <-- Alignments: 1st column left, 2nd middle and 3rd right, with vertical lines in between
%   {Beam Parameter} &{Unit}& \multicolumn{3}{c}{L1}&\multicolumn{3}{c}{L2}& \\
 &&WP1&WP2&WP3\\
      \hline
Charge&$Q (pC)$&30&30&200\\
Energy&$E (GeV)$&0.5&0.5&1.0\\
Hor rms beam size&$\sigma_x (\mu m)$&2.&1.&30.\\
StDev of $\sigma_x $&$(\mu m)$&0.03&0.1&10.\\
StDev of Hor Centroid position&$\sigma_{C_x} (\mu m)$&0.5&0.4&10.\\
Ver rms beam size&$\sigma_y (\mu m)$&1.&1.&40.\\
StDev of $\sigma_y $&$(\mu m)$&0.01&0.02&5.\\
StDev of Ver Centroid position&$\sigma_{C_y} (\mu m)$&0.3&0.3&30.\\
Rms Energy spread&$\sigma_{\delta} (\%)$&0.07&0.08&0.14\\
     \end{tabular}
  \end{center}
\end{table}
\section{Conclusions}
The X-band Linac for the EUPRAXIA@SPARC\_LAB facility is under design and at this stage it has been focused on three different working points to explore the line acceptance and robustness. The beam dynamics studies have been presented for the nominal cases and, as first test,  misalignments for the active elements (RF and magnetic) have been considered and studied to have an indication on the required tolerances and machine operation scenario. The very next steps will include RF phase and amplitude jitters, and Photocathode laser energy and pointing jitters as well. The space charge effects and possible disomogeneities of the permanent focusing quadrupoles will also be investigated in the strong focusing sections before the injection in the plasma.
\section*{Acknowledgments}
This work was supported by the European Union's Horizon 2020 research and innovation programme under grant agreement No. 653782.
\section*{References}
\bibliographystyle{elsarticle-num}
\bibliography{CRibibfile}

\begin{thebibliography}{10}
\expandafter\ifx\csname url\endcsname\relax
  \def\url#1{\texttt{#1}}\fi
\expandafter\ifx\csname urlprefix\endcsname\relax\def\urlprefix{URL }\fi
\expandafter\ifx\csname href\endcsname\relax
  \def\href#1#2{#2} \def\path#1{#1}\fi

\bibitem{Eupraxia_EU}
P.~Walker, et~al., {HORIZON} 2020 {E}u{PRAXIA} {D}esign {S}tudy, in: Proc. of
  IPAC2017, TUOBB3, Copenhagen, Denmark, 2017.

\bibitem{Eupraxia_SPARC_LAB}
M.~Ferrario, et~al., {E}u{PRAXIA}@{SPARC\_LAB}: design study towards a compact
  {FEL} facility at {LNF}, these Proceedings.

\bibitem{SPARC_proj}
D.~Alesini, et~al., The {SPARC} project: {A} high-brightness electron beam
  source at {LNF} to drive a {SASE-FEL} experiment, Nuclear Instruments and
  Methods in Physics Research, Section A: Accelerators, Spectrometers,
  Detectors and Associated Equipment 507~(1--2) (2003) 345--349.
\newblock \href {http://dx.doi.org/10.1016/S0168-9002(03)00943-4}
  {\path{doi:10.1016/S0168-9002(03)00943-4}}.

\bibitem{Anna_PI}
A.~Giribono, et~al., {E}u{PRAXIA}@{SPARC\_LAB}: the {H}igh {B}rightness {RF}
  {P}hotoinjector {L}ayout {P}roposal, theese Proceedings.

\bibitem{ChiadroniCuba2017}
E.~Chiadroni, D.~Alesini, M.~Anania, A.~Bacci, M.~Bellaveglia, A.~Biagioni,
  F.~Bisesto, F.~Cardelli, G.~Castorina, A.~Cianchi, M.~Croia, A.~Gallo, D.~D.
  Giovenale, G.~D. Pirro, M.~Ferrario, F.~Filippi, A.~Giribono, A.~Marocchino,
  A.~Mostacci, M.~Petrarca, L.~Piersanti, S.~Pioli, R.~Pompili, S.~Romeo,
  A.~Rossi, J.~Scifo, V.~Shpakov, B.~Spataro, A.~Stella, C.~Vaccarezza,
  F.~Villa, Beam manipulation for resonant plasma wakefield acceleration,
  Nuclear Instruments and Methods in Physics Research A 865 (2017) 139.

\bibitem{beam_driven}
A.~Marocchino, E.~Chiadroni, M.~Ferrario, F.~Mira, A.~R. Rossi, {D}esign of
  high brightness {P}lasma {W}akefield {A}cceleration experiment at
  {SPARC\_LAB} test facility with particle-in-cell simulations, theese
  Proceedings.

\bibitem{stefanoromeo}
S.~Romeo, E.~Chiadroni, M.~Croia, M.~Ferrario, A.~Giribono, A.~Marocchino,
  F.~Mira, R.~Pompili, A.~Rossi, C.~Vaccarezza, Simulation design for
  forthcoming high quality plasma wakefield acceleration experiment in linear
  regime at {SPARC\_LAB}, theese Proceedings.

\bibitem{Andrea_LW}
A.~Rossi, V.~Petrillo, A.~Bacci, M.~Ferrario, A.~Giribono, A.~Marocchino, M.~R.
  Conti, L.~Serafini, C.~Vaccarezza, {Plasma} boosted electron beams for
  driving {Free} {Electron} {Lasers}, theese Proceedings.

\bibitem{Vittoria}
V.Petrillo, G.~Dattoli, A.Petralia, F.~Vill, M.~Ferrario, A.~Bacci, L.~Serafin,
  E.~Chiadroni, A.~Giribono, A.~Marocchino, M.~R. Conti, A.~Rossi,
  C.~Vaccarezza, {Free} {Electron} {Laser} in the water window with plasma
  driven electron beams, theese Proceedings.

\bibitem{M_Diomede}
M.~Diomede, D.~Alesini, M.~Bellaveglia, B.~Buonomo, F.~Cardelli, N.~C.
  Lasheras, E.~Chiadroni, G.~D. Pirro, M.~Ferrario, A.~Gallo, A.~Ghigo,
  A.~Giribono, A.~Grudiev, L.~Piersanti, B.~Spataro, C.~Vaccarezza, W.~Wuensch,
  Preliminary {RF} design of an {X}-band linac for the
  {E}u{PRAXIA}@{SPARC\_LAB} project, theese Proceedings.

\bibitem{Tstep}
L.~Young, Tstep:an electron linac design code.

\bibitem{Elegant}
M.~Borland, Elegant: A flexible sdds-compliant code for accelerator simulation,
  Advanced Photon Source LS-287.

\bibitem{K_Bane}
K.~Bane, Short-range dipole wakefields in accelerating structures for nlc,
  SLAC-PUB-9663.

\end{thebibliography}
\end{document}